\documentstyle[12pt,epsf]{article}

\newcommand{\rsection}[1]{\section{#1}\setcounter{equation}{0}}
\newcommand{\rcite}[1]{{\cite{#1}}}
\newcommand{\rref}[1]{{(\ref{#1})}}
\newcommand{\tref}[1]{{\ref{#1}}}
\newcommand{\rlabel}[1]{{\label{#1}}}
\newcommand{\rbibitem}[1]{\bibitem{#1}}
\newcommand{\be}{\begin{equation}}
\newcommand{\ee}{\end{equation}}
\newcommand{\ba}{\begin{eqnarray}}
\newcommand{\ea}{\end{eqnarray}}
\newcommand{\tr}{{\rm tr}}
\newcommand{\disp}{\displaystyle}
\begin{document}
\begin{titlepage}
\begin{flushright}
NORDITA - 95/12 N,P\\
hep-ph/9502393
\end{flushright}
\vspace{2cm}
\begin{center}
{\Large\bf Chiral Perturbation Theory\footnote{
Invited Talk at
the International Workshop on Nuclear \& Particle Physics ``Chiral Dynamics in
Hadrons \& Nuclei,'' Feb 6 $\sim$ Feb 10, 1995, Seoul, Korea. To be published
in the proceedings.}
}\\
\vfill
{\bf Johan Bijnens}\\[0.5cm]
NORDITA, Blegdamsvej 17\\
DK-2100 Copenhagen \O, Denmark
\end{center}
\vfill
\begin{abstract}
A short overview of the current state of Chiral Perturbation Theory
is given. This includes a description of the basic assumptions,
the usefulness of the external field method is emphasized using
a simple lowest order example. Then at next-to-leading
order the determination of the parameters is discussed. We also
present the status of calculations at ${\cal O}(p^6)$.
Finally I present the extension into 3 directions: estimates
of the free parameters, inclusion of nonleptonic weak and electromagnetic
interactions, and inclusion of non-Goldstone fields in the chiral
Lagrangian.
\end{abstract}
\vfill
February 1995
\end{titlepage}
\begin{center}
{\Large\bf
     Chiral Perturbation Theory}\\
\vskip 1.0cm
                  {\bf JOHAN BIJNENS}\\
    {\it  NORDITA, Blegdamsvej 17} \\
         {\it  DK 2100 Copenhagen \o, Denmark}
\end{center}
\vskip 1.5cm

\begin{abstract}
A short overview of the current state of Chiral Perturbation Theory
is given. This includes a description of the basic assumptions,
the usefulness of the external field method is emphasized using
a simple lowest order example. Then at next-to-leading
order the determination of the parameters is discussed. We also
present the status of calculations at ${\cal O}(p^6)$.
Finally I present the extension into 3 directions: estimates
of the free parameters, inclusion of nonleptonic weak and electromagnetic
interactions, and inclusion of non-Goldstone fields in the chiral
Lagrangian.
\end{abstract}

\rsection{Introduction}
Chiral Perturbation Theory (CHPT) has become since the seminal work of
Gasser and Leutwyler \rcite{GL1,GL2} a more and more popular method to
treat hadronic phenomena at low energy. In this talk I will give a review
of the present situation and comment about some points that have
recently been the subject of discussions.
More extensive reviews have appeared recently. Here Ref. \rcite{eduardo1}
is mainly concerned with the purely mesonic sector and CP violation, Ref.
\rcite{ecker} discusses purely mesonic processes and those with one baryon
line. Both of
these also talk about the
3 light flavour case.
Ref. \rcite{bernard1} concentrates on processes involving
one or more nucleons in the two light flavour case. A more introductory one
is Ref. \rcite{Pich}.

A more pedestrian introduction can be found in the recent book by Donoghue,
Golowich and Holstein \rcite{Donoghue}. Very up-to-date reviews of various
mesonic processes can be found in the DA$\Phi$NE handbook
\rcite{dafne}. This contains
amongst others a short introduction to CHPT\rcite{Bijnens1}
and an overview of semileptonic Kaon decays\rcite{Bijnens2}. There is also
the proceedings of the MIT workshop on Chiral Dynamics in july 1994\rcite{MIT}.

One of the points not discussed in this talk is the inclusion of
of heavy quarks and heavy quark
symmetry. A point of entry in the literature is the Physics Reports
by Neubert \rcite{Neubert}. Also not discussed are applications outside
hadronic physics.

This talk is organized as follows. In Sect. \tref{basics} I discuss the
principles underlying the method. The lowest order mesonic Lagrangian,
a simple example of an amplitude including its off-shell definition
and a discussion of low-energy theorems is given in the next section.
In section \tref{p4} I treat the next order in the chiral expansion
and a determination of its parameters including quark masses.
This includes an example of the use of dispersion relations.
Sect. \tref{p6} reviews the present status of order $p^6$.
The next section discusses some attempts at estimating the numerous
free parameters from underlying models. The last two sections
discuss extensions of the basic mesonic theory into two directions,
inclusion of nonleptonic weak and electromagnetic interactions, Sect.
\tref{nonleptonic}, and non-Goldstone fields, Sect. \tref{NGB}.
The last section summarizes the present situation.

\rsection{Basics}
\rlabel{basics}
First we have to define what precisely is Chiral Perturbation Theory.
A loose way to phrase it is:\\
{\em Chiral Perturbation Theory is for the Goldstone Theorem what
Clebsch-Gordan
coefficients are for the Wicker-Eckart Theorem.}\\
Alternatively:\\
{\em Chiral Perturbation Theory is a systematic method to use
a spontaneously broken symmetry}.\\
The main underlying assumption has been phrased as a theorem by
Weinberg\rcite{weinberg1}:\\
{\em The most general solution of causality, unitarity and symmetry
in quantum field theory is given
by the most general symmetric Lagrangian.}\\
This includes all the loop diagrams generated by this Lagrangian.
In the case of Goldstone Bosons as the only relevant degree of freedom
this has in fact been proven recently\rcite{leutwyler1,weinberg2}.
In the first reference it is also proven that the Lagrangian can be local.
The exception to the theorem mentioned above is the possible occurrence
of Wess-Zumino type terms. These change the Lagrangian by a total derivative
and thus leave the action invariant. A very important ingredient of this
proof was Lorentz invariance. Relaxation of this requirement, as is the case
if we want to write an effective theory for mesons only in the presence
of a baryonic background or for spin waves in a solid there are
more terms possible\rcite{leutwyler2}.

{}From the above it is obvious that the method has a wide range of
applicability
whenever there is a symmetry spontaneously broken. Areas which are not
discussed
here are studies of the symmetry breaking sector in the standard model
and applications to solid state physics. I will
concentrate on the realm of hadronic physics of the 3 light quarks.
For extensions including heavy quarks see ref. \rcite{Neubert}.

The symmetry that is spontaneously broken is the flavour symmetry
of the up, down and strange quarks. If the masses are zero the QCD Lagrangian
does not contain any terms coupling the left and right handed chirality.
The classical symmetry then is the one where the left and right handed
quarks transform separately. For $g_{L(R)} \in U(3)_{L(R)}$, the quarks
transform as
\be
\rlabel{qtransf}
q\to g_L q_L + g_R q_R\qquad{\rm with}\qquad
q = \left(\begin{array}{c}u\\d\\s\end{array}\right)
\ee
and $q_{L(R)} = 1/2(1+(-)\gamma_5) q$. We thus have a $U(3)_L\times U(3)_R$
flavour symmetry. The axial combination of the singlet factors,
$U(1)_A = U(1)_{R-L}$, is coupled via the anomaly to gluons and is
not conserved. Some of the other generators are via the anomaly coupled
to the photon and the weak vector bosons, $W$ and $Z$, but these effects
are proportional to the electromagnetic or weak coupling constant and can
thus be treated perturbatively. These parts of the symmetry thus remain
usable in CHPT. The usable symmetry for CHPT is thus
\be
\rlabel{G}
G = SU(3)_L \times SU(3)_R \times U(1)_V\ .
\ee
This symmetry is not seen as an explicit symmetry in nature. In that case
there would have to be parity doublets for every massive observed hadronic
state. This is definitely not the case in the observed hadronic
spectrum. There are candidates for the Goldstone degrees of freedom
that a spontaneously broken symmetry would require, the lightest pseudoscalar
$SU(3)$ octet consisting of $\pi$, $K$ and $\eta$. In addition
all indications are that there is a nonzero order parameter that breaks
the symmetry $G$ spontaneously, $\langle \overline{q}_Lq_R + \overline{q}_R q_L
\rangle \ne 0$. The symmetry observed (at least approximately) in the
hadronic spectrum is
\be
H = SU(3)_{L+R} \times U(1)_V\ .
\ee
The $U(1)_V$ plays only a minor role in CHPT. The mesons do not transform
under it and in the baryon sector its main role is the conservation
of the number of baryons.

The Goldstone theorem requires massless degrees of freedom living in the
broken part of the group. In this case this is
$G/H = SU(3)$. There are thus 8 Goldstone Bosons. Another consequence
of the Goldstone theorem is that at low energies interactions between
these Goldstone Bosons are suppressed. The interaction contains
at least two powers of momenta.
This fact allows to replace the usual expansion in a (small) coupling constant
by an expansion in the number of derivatives. This was done in general in
\rcite{weinberg1}. I will demonstrate it here on the example of $\pi\pi$
scattering.
\begin{figure}\begin{center}\leavevmode
\epsfbox{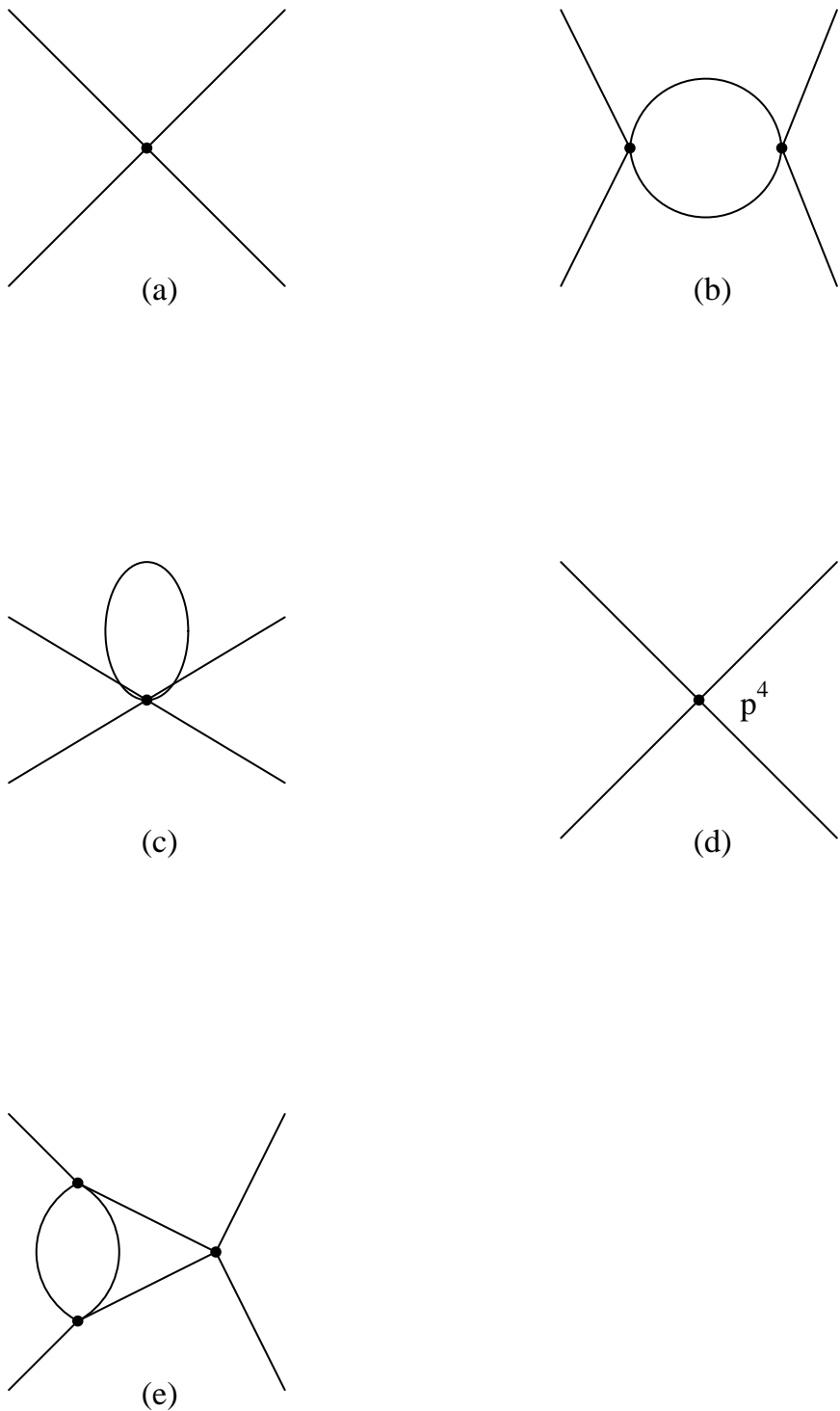}
\end{center}
\caption{The diagrams at lowest order (a), next-to-leading order (b-d) and
an example of a $p^6$ diagram (e) for $\pi\pi$ scattering}
\rlabel{fig1}
\end{figure}
This is dimensional counting.
The diagram in Fig. \tref{fig1}a is the lowest order diagram. The vertex
contains two derivatives and this contributes to the amplitude at order $p^2$.
The diagram in Fig. \tref{fig1}b contains two propagators, $p^{-4}$, and two
vertices with two derivatives, $p^4$, and a loop integration, $d^4p$. Putting
all the dimensions together leads to an amplitude of order $p^4$. Similarly
for the diagram in Fig. \tref{fig1}c. Here there is one vertex, one propagator
and one loop integration leading again to an amplitude of order $p^4$.
Finally there is the tree level contribution with a vertex with 4 derivatives.
This is immediately order $p^4$. Then the infinities arising in the loop
diagrams can be systematically absorbed in the free parameters of the $p^4$
Lagrangian \rcite{weinberg1,GL1,GL2}. This counting can be easily generalized
to all orders. As an example, the two-loop diagram of Fig. \tref{fig1}e
is of order $p^6$. The lowest order amplitude for this process was determined
by Weinberg\rcite{Weinberg3}. The $p^4$ was worked out in Ref. \rcite{GL1}.
Work on the $p^6$ amplitude is in progress\rcite{bijnens3}.

One more ingredient has to be added. This is the method of using a generating
functional using the external field formulation as introduced by Gasser and
Leutwyler\rcite{GL1,GL2}. This method has two advantages. It is obviously
independent of the parametrization chosen for $G/H$ and allows thus for a
well defined definition of off-shell amplitudes. As a simple example,
see Sect. \tref{p2}, the amplitude for the decay $K\to\pi W$ where the $W$
is the weak vector boson can be simply extrapolated off-shell via the
Green function $\langle 0 | T\left( a_\mu^K v_\alpha^W a_\nu^\pi
\right)\rangle$. Here the axial currents couple to the $K$ and $\pi$,
respectively, and the vector coupling couples to the $W$. This is a $SU(3)$
rotation of the pion vector form factor.

The second advantage of using this method is that the connection with QCD
becomes clearer. The generating functional in terms of external vector
($v_\mu$),
axial-vector ($a_\mu$), scalar ($s$) and pseudoscalar ($p$) external fields
at low-energies is given by
\ba
e^{\disp\left(i\Gamma\left(v_\mu,a_\mu,s,p\right)\right)} &\equiv&
\frac{1}{Z}\int[dq][dG]e^{\disp i\int d^4x\left({\cal L}^0_{QCD}
+\overline{q}\gamma^\mu(v_\mu+a_\mu\gamma_5)q-
\overline{q}(s-ip\gamma_5)q\right)}
\nonumber\\
&\approx&\frac{1}{Z}\int[dU]e^{
\disp i\int d^4x{\cal L}_{CHPT}\left(U,v_\mu,a_\mu,s,p\right)}
\ .
\ea
The first line is the definition of the generating functional and the second
line the approximation valid at low energies.

\rsection{Lowest Order}
\rlabel{p2}

In the case presented here the most convenient parametrization of the
Goldstone boson space, $G/H = SU(3)$ is the exponential parametrization:
\be
U = \exp\left(\frac{i\sqrt{2}}{F}\Phi\right)
\qquad {\rm with}\qquad
\Phi=\frac{1}{\sqrt{2}}\pi^a\lambda^a=
\left(\begin{array}{ccc}
\frac{\pi^0}{\sqrt{2}}+\frac{\eta}{\sqrt{6}} & \pi^+ & K^+\\
\pi^- & \frac{\pi^0}{\sqrt{2}}+\frac{\eta}{\sqrt{6}} & K^0\\
K^- &\overline{K^0} & \frac{-2\eta}{\sqrt{6}}\end{array}\right)\ .
\ee
The transformation under the chiral transformation $g_L \times g_R \in
SU(3)_L\times SU(3)_R$ is
\ba
U&\to&g_R U g_L^\dagger\nonumber\\
r_\mu = v_\mu+a_\mu &\to& g_R r_\mu g_R^\dagger + i g_R\partial_\mu g_R^\dagger
\nonumber\\
l_\mu = v_\mu-a_\mu &\to& g_L l_\mu g_L^\dagger + i g_L\partial_\mu g_L^\dagger
\nonumber\\
s+ip &\to& g_R \left( s+ip\right) g_L^\dagger\nonumber\\
q &\to& g_R q_R + g_L q_L\ .
\ea
We then start constructing an effective Lagrangian in terms of derivatives and
external fields.
There is no term without derivatives and external fields. The only possibility
would be $\tr\left( U U^\dagger\right) = 3$
and that is just a constant. Therefore the
lowest order Lagrangian starts at the two derivative level:
\be
{\cal L}_2 = \frac{F^2}{4} \tr \left( D_\mu U D^\mu U^\dagger
+\chi U^\dagger + U \chi^\dagger \right) \ ,
\ee
with $\chi = 2B_0 (s+ip)$ and $D_\mu U = \partial_\mu U -i r_\mu U + iU l_\mu$.
The tree level diagrams resulting from this simple Lagrangian reproduce
most of the mesonic current algebra results. There are two free parameters
at this order, $F$ and $B_0$.

Let us now turn to the example of a well defined off-shell amplitude.
A similar discussion in a different process can be found in
Ref.~\rcite{Vesteinn}.
\begin{figure}\begin{center}\leavevmode
\epsfxsize=12cm\epsfbox{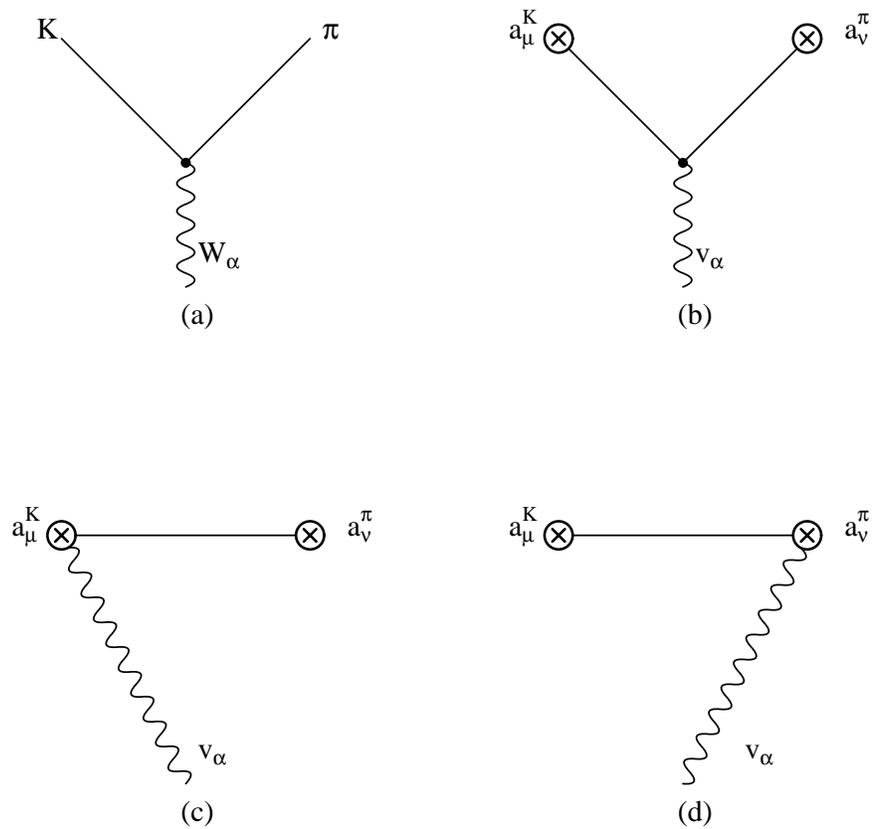}
\end{center}
\caption{(a) The diagram for $K\to\pi W$ in the standard Feynman Diagram
approach; (b-d) the diagrams in the external field method.}
\rlabel{fig2}
\end{figure}
The amplitude for the process $K(p_K)\to \pi(p_\pi) W$
from the diagram in Fig.~\tref{fig2}a
is given by
\be
\rlabel{simple}
A = \frac{i}{\sqrt{2}}\left(p_\pi + p_K\right)_\alpha W^\alpha\ .
\ee
In the limit of equal quark masses this amplitude satisfies the correct
behaviour only on-shell, i.e. $p_\pi^2 = p_K^2 = m_\pi^2$. Off-shell it does
not vanish but is proportional to $p_\pi^2-p_K^2$ when $W^\alpha$ is replaced
by
$(p_K-p_\pi)^\alpha$.
In sharp contrast the amplitude calculated in the external field formalism
corresponds to
\be
\rlabel{extfield}
i^3\langle T\left(a_\mu^K v_\alpha a_\nu^\pi \right)\rangle =
i\sqrt(2)F^2\left\{
\frac{\left(p_\pi+p_K\right)_\alpha p_{\pi\nu} p_{K\mu}}{\left(p_\pi^2-m_\pi^2
\right)\left(p_K^2-m_K^2\right)}
-\frac{g_{\mu\alpha}p_{\pi\nu}}{p_\pi^2-m_\pi^2}
-\frac{g_{\nu\alpha}p_{K\mu}}{p_K^2-m_K^2}
\right\}\ .
\ee
The diagrams are depicted in Fig.~\tref{fig2}(b-d). The circled crosses
are insertions of the axial currents.
This amplitude is well defined off-shell and satisfies the correct
Ward-Identity. If the external legs are reduced and after going on-shell it
agrees with Eq. \rref{simple}. However, after making the replacement of
$v^\alpha$  by $(p_K-p_\pi)^\alpha$ we obtain the correct Ward identity for
all values of masses and momenta. Similarly only the amplitudes which are
defined using this method but with $\partial_\mu a^\mu$ satisfy the
off-shell current algebra relations and not the on-shell amplitudes
like in Eq.~\rref{simple} that are extrapolated off-shell.

Let me close this session with a few simple remarks about low-energy theorems.
There has been some confusion, see e.g. the discussions in \rcite{MIT}.
The underlying problem is that there are different types of low-energy
theorems and one should carefully distinguish between them. Three common
types are
\begin{enumerate}
\item Low low-energy theorems: These are valid for photon radiation in the
limit
of vanishing photon mass as derived by Low. They relate the process with a
soft-photon to the one without. This is an expansion in $E_\gamma$.
\item Chiral low-energy theorem: these are CHPT predictions to a given order
in the chiral expansion. They relate different processes to each other in
terms of the CHPT parameters to any order. These require $m_\pi$ and external
pion momenta small. If done correctly the PCAC relations correspond exactly
to these.
\item Multipole low-energy theorems: this is an expansion of the amplitudes
in multipoles and then only keeping the lowest ones.
In addition one often expands also in other
kinematical variables and this typically requires $|E_\pi - m_\pi|<<m_\pi$.
Their regime of validity thus requires a small kinetic energy.
\end{enumerate}
In most case of interest several of these apply. E.g., in $\gamma N\to \pi N$
both 2. and 3. apply and the amplitudes of Bernard et al. \rcite{kaiser}
satisfy the multipole expansion if the expansion in $(E_\pi-m_\pi)$ is done.
They do also show that this expansion has a very small domain of validity.

\rsection{Next-to-Leading Order and the Values of its Parameters}
\rlabel{p4}

At the next-to-leading order there are 12 terms plus the Wess-Zumino term.
The explicit form of the
Lagrangian can be found in Refs. \rcite{GL2} to \rcite{Bijnens1}.
The Wess-Zumino term describes the anomaly and has a fixed coefficient.
Of the remaining 12 terms two are not measurable. They correspond to specific
choices of the external field renormalization in QCD. So we
have 10 new parameters that need to be determined experimentally.

In addition there are ambiguities in the effective theory itself in
identifying the quark masses. The reason is that $\chi' = \left(\chi^\dagger
\right)^{-1}\det\chi$ has the same transformation properties as $\chi$ under
the chiral group.
Replacing $\chi$ by $\chi+\beta\chi^\prime$ corresponds to a shift in the
values of $L_6$, $L_7$ and $L_8$ and $m_u \to m_u +\alpha m_d m_s$. This is
known as the Kaplan Manohar ambiguity\rcite{kaplan}. This problem has two
solutions:
\begin{enumerate}
\item $\chi^\prime$ transforms differently under $U(1)_A$ then
$\chi$\rcite{DW}.
\item go to QCD directly. This is equivalent to calculating the relevant
coefficients $L_i$ thus fixing the 'shift'.
\end{enumerate}
The latter approach has been done in the QCD sum rule and lattice determination
of quark masses:
\be
2\hat m=(m_u+m_d)(1~{\rm GeV}^2) = 12\pm2.5~{\rm MeV}
\ee
as derived in ref. \rcite{BPR} and
\be
m_s(1~{\rm GeV}^3) = 175\pm25~{\rm MeV}
\ee
from Refs. \rcite{ms}. In \rcite{BPR} the quark vacuum expectation value was
also determined:
\be
\frac{1}{2}\langle\overline{u}u+\overline{d}d\rangle =
-(0.013\pm0.003)~{\rm GeV}^3 = -(235~{\rm MeV})^3\ .
\ee
This leads to a large value for $B_0$ and a small ($\approx$3.5\%) correction
to the Gell-Mann-Oakes-Renner relation. We can add in addition the relation
\rcite{GL2}
\be
\frac{m_d-m_u}{m_d+m_u}=
\frac{m_\pi^2}{m_K^2}
\frac{\left(m_{K^0}^2-m_{K^+}^2\right)_{QCD}}{m_K^2-m_\pi^2}
\frac{m_s^2\hat{m}^2}{4\hat{m}^2}
\ee
together with determination of the electromagnetic part of the mass
difference\rcite{mk} to obtain
\be
\frac{m_u}{m_d}=0.44\pm0.22\ .
\ee
Notice that the numbers above lead to $m_s/\hat{m}=29\pm7$, very close to the
current algebra values.

Using these quark mass values the values of the $L_i$ can then be determined
\rcite{GL2,Bijnens2}.
These are in Table \tref{table1} where I have also listed the source
of the experimental information used.
\begin{table}
\begin{center}
\begin{tabular}{|c|c|c|}
\hline
$L_i$ & Value $\cdot 10^3$ & Input\\
\hline
1 & $0.4\pm0.3 $& $K_{e4}$ and $\pi\pi\to\pi\pi$\\
2 & $1.35\pm0.3 $& $K_{e4}$ and $\pi\pi\to\pi\pi$\\
3 & $-3.5\pm1.1$ & $K_{e4}$ and $\pi\pi\to\pi\pi$\\
4&$-0.3\pm0.5$&$1/N_c$ arguments\\
5&$1.4\pm0.5$&$F_K/F_\pi$\\
6&$-0.2\pm0.3$&$1/N_c$ arguments\\
7&$-0.4\pm0.2$&GMO, $L_5$, $L_8$\\
8&$0.9\pm0.3$&$m_{K^0}-m_{K^+}$, $L_5$, baryon mass ratios\\
9&$6.9\pm0.7$&pion electromagnetic charge radius\\
10&$-5.5\pm0.7$&$\pi\to e\nu\gamma$\\
\hline
\end{tabular}
\end{center}
\caption{The values of the $L_i$ coefficients and the input used
to determine them, they are quoted at a scale $\mu=m_\rho$.}
\rlabel{table1}
\end{table}

Now the first three are from $K\to\pi\pi e\nu$\rcite{BCG}. In amplitudes
they are determined from the formfactor. As an
example I quote the $s$ wave one at threshold.
The lowest order calculation gives $f_S(0)=3.74$ and the experimental
determination was $f_S(0)=5.59\pm0.14$. So there is a 50\% correction
going to higher order. The question is can we now trust a
next-to-leading order calculation. We can answer part of this since the
sources of large higher order corrections are known. We can then
use the strategy (see \rcite{dispersion}) of using dispersion
relations and determining the subtraction constants using CHPT to
estimate the higher orders. This was done in Ref. \rcite{BCG}
for the first three coefficients.
We obtained $L_{1(2)}=0.60(1.5)\cdot 10^{-3}$ at the one-loop accuracy
and $L_{1(2)}=0.37(1.35)\cdot10^{-3}$ estimating the higher orders with
dispersion relations. So the size of the higher orders seems under control
for these processes.

\rsection{Order $p^6$}
\rlabel{p6}

The situation at order $p^6$ is somewhat less complete. There exists
a classification of all terms in the Lagrangian at this order\rcite{Scherer}.
For the sector including an odd number of Levi-Civita tensors
($\varepsilon^{\mu\nu\alpha\beta}$), a lot of calculations exist and the
general infinity structure is known\rcite{Bijnens3}.
In this case $p^6$ is the next-to-leading
order. Some two-loop calculations also exist. In particular the $p^6$
correction
to $\gamma\gamma\to\pi^0\pi^0$ is known\rcite{BGS} and several more
calculations are in progress.
\begin{figure}\begin{center}\leavevmode
\epsfxsize=12cm\epsfbox{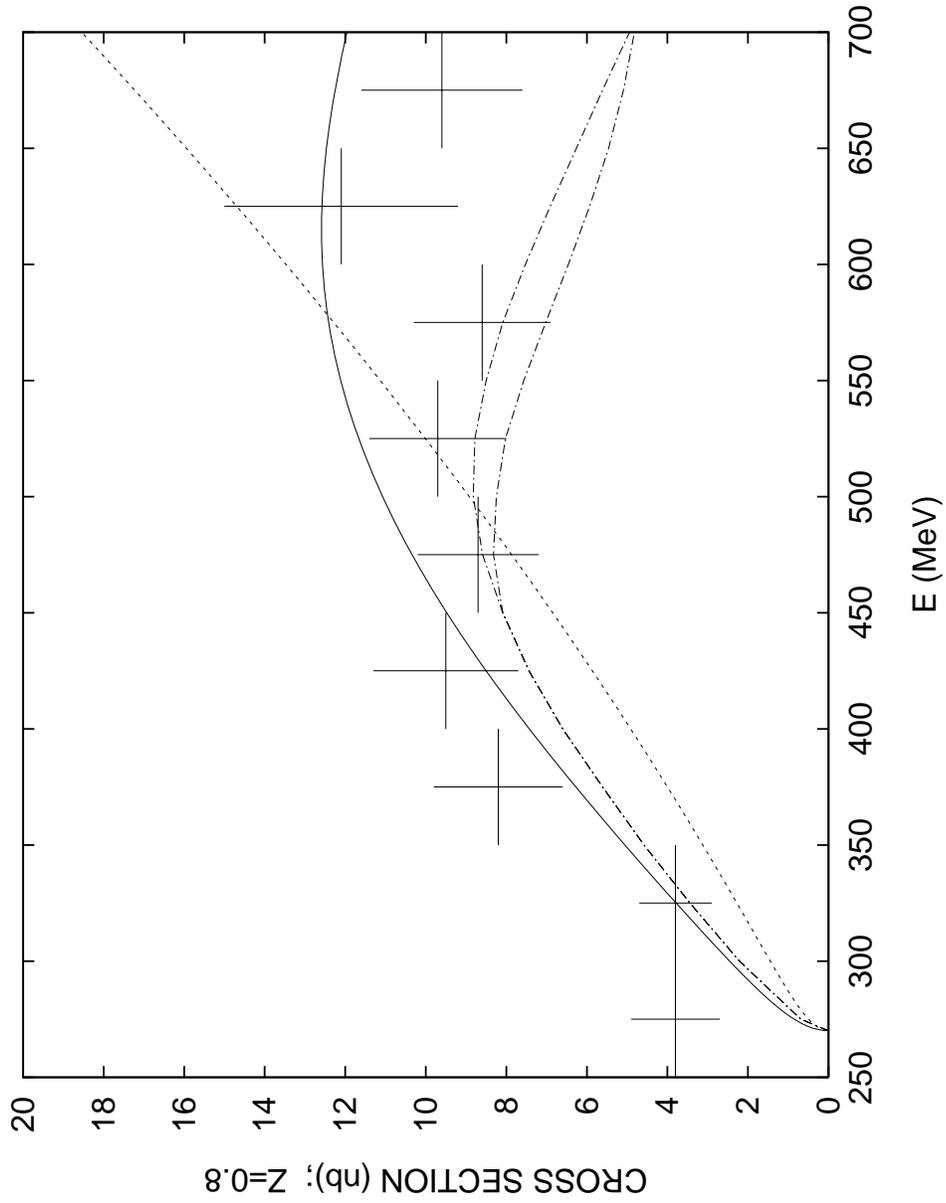}
\end{center}
\caption{Experimental results (crosses), $p^4$ calculation (dashed), $p^6$
calculation (full) and dispersive estimates (dash-dotted), taken from
Ref.~\protect{\rcite{BGS}}.}
\rlabel{fig3}
\end{figure}
In Fig. \tref{fig3} I have shown the effect of the one-loop calculation for
$\gamma\gamma\to\pi^0\pi^0$. This was in fact a parameter free prediction.
The dispersive calculation and the $p^6$ calculation are in impressive
agreement
with each other and with the data.

In general calculations at this order are technically very demanding and still
contain a reasonably large number of free parameters. It thus becomes
necessary to estimate those coefficients from other sources.

\rsection{Estimates of Parameters}
\rlabel{param}

The first attempts at estimating the $L_i$ from underlying physics
arguments were done in Refs. \rcite{Ecker1,Ecker2,Donoghue4}
and in Ref. \rcite{Bijnens4} for the anomalous sector. The basic
idea is that formfactors are dominated by resonance exchange.
E.g., the pion electromagnetic form factor is dominated
by $\rho$ exchange, $F_\pi(q^2)\approx 1/\left(1-q^2/m_\rho^2\right)$,
leading to the prediction $L_9 = F_V G_V/(2m_\rho^2)$.
This type of estimates was used in the calculation in Ref.~\rcite{BGS}.
In the anomalous sector there is a problem with trying to implement
full meson dominance\rcite{Bijnens5} but one can still
estimate the order $p^6$ parameters.

One can also use constraints from high energy behaviour\rcite{Ecker2}.

The third avenue is to calculate them from models intermediate between
QCD and CHPT. A most prominent example is the calculation in the ENJL model.
See Ref.~\rcite{Bijnens6} and references therein. This model in fact
leads to most of the meson dominance relations obtained using the first method.

\rsection{Inclusion of Nonleptonic Weak and Electromagnetic Interactions}
\rlabel{nonleptonic}

Here we need to construct terms in the effective Lagrangian that corresponds
to the nonleptonic part of the electromagnetic and weak interaction.
Let me concentrate on the electromagnetic example.
The underlying effective action is
\be
\rlabel{heff}
{\cal H}_{\rm eff} =
\int d^4q \frac{-i e^2}{q^2}\left(g_{\mu\nu}-\frac{q_\mu q_\nu}{q^2}\right)
J_{\rm em}^\mu J_{\rm em}^\nu\ .
\ee
This effective Hamiltonian transforms under $SU(3)_L\times SU(3)_R$ as
$\left(8_L +8_R\right)^2$. So we now need to construct terms
using the CHPT external fields and degrees of freedom, $U$,
that transform in this fashion. This we
do via introducing {\em spurion} fields. These fields are dummy fields that
are added to the terms like Eq.~\rref{heff} to make them singlets under
the chiral group. This procedure is similar to the one used for inclusion of
the quark masses. The term $\overline{q}_iq_i$ is made invariant by introducing
the scalar field $s$. $-s^{ij}\overline{q}_iq_j$ has singlet properties under
the chiral group. In the chiral Lagrangian we then include the field $s$
via
$\frac{2 B_0 F^2}{4}\tr\left(sU+U^\dagger s\right)$. The quark
masses are then included later via $s={\rm diag}\left(m_u,m_d,m_s\right)$.

The same thing can now be done for the nonleptonic Lagrangians. Eq. \rref{heff}
contains a term $\chi^{ij}_L\chi^{kl}_R
\quad\overline{q}_{iL}\gamma_\mu q_{jL}\quad
\overline{q}_{kR}\gamma^\mu q_{lR}$. This is made invariant by making $\chi_L$
and $\chi_R$ transform as left and right handed octets under the chiral group.
There are then two terms that can be constructed at lowest order:
\be
c\;\tr \left(\chi_R U \chi_L U^\dagger\right) + d\;\tr
\left(\chi_L^2+\chi_R^2\right)\ .
\ee
Putting in the right values for $\chi^{ij}_L$ and $\chi^{kl}_R$ this term
is then responsible for the $\pi^+-\pi^0$ mass difference.
At higher orders one can then similarly construct all terms.

Unfortunately this leads to very large numbers of terms at next-to-leading(NLO)
order. For the nonleptonic electromagnetic case these have been classified by
Urech\rcite{Urech}. As shown above at lowest order there are 2, one of
which is a pure counterterm. At NLO there are 15. Here in fact there
are large corrections expected\rcite{mk}.

In the weak nonleptonic sector the terms and the associated infinity structure
has been classified by Kambor et al. \rcite{Kambor}. Here there is one
parameter
at leading order each for octet and 27 (or $\Delta I=3/2$)
 transitions but at NLO there are
48 parameters in the octet case and 34 for the 27 case.
Here it thus becomes very important to be able to estimate these from other
sources. The main problem is that, as in Eq.~\rref{heff} there is an
integration
over the momentum of an external gauge field.
This problem thus involves the strong
interaction at {\em all} scales. The main attempts are done
using factorization, quark models \rcite{pich}, ENJL\rcite{Bijnens8}
and various sum rules\rcite{dispersion}. See also the references in these
papers.

\rsection{Inclusion of non Goldstone Boson Fields}
\rlabel{NGB}

{\bf a) Vector Mesons:} In this sector we loose pure CHPT
power counting since due to the
diagram of Fig.~\tref{fig4} whenever there is a vector meson on-shell
present, it always involves large momenta for the pseudoscalars in the
intermediate state.
\begin{figure}\begin{center}\leavevmode
\epsfxsize=8cm\epsfbox{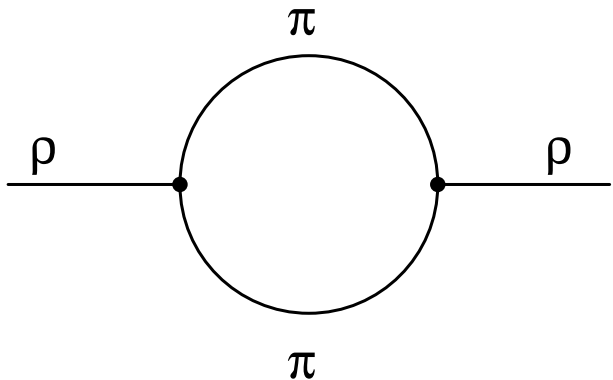}
\end{center}
\caption{A diagram creating problems for power counting in CHPT including
on-shell vector mesons.}
\rlabel{fig4}
\end{figure}
The problem is then that we need counterterms involving
pions to all orders. This does not invalidate the discussion in
Sect.~\tref{param}, there the vector meson was at low momentum.
In general the use of effective Lagrangians for meson fields can still
be useful (see E.g. the talks by Ko and Pisarski in these proceedings).
The choice of interpolating field for the $\rho$ is free. This leads
to the different representations for the vector field:
\begin{enumerate}
\item The Gauged Yang-Mills version
\item Hidden Local gauge Symmetry version
\item Using the standard CCWZ mechanism (see in \rcite{Donoghue})
\item The anti-symmetric tensor field representation, used in \rcite{Ecker1}
\end{enumerate}
These are all equivalent but some choices of parameters look nice in one
version
and ugly in another one. As an example, the vector meson decay vertex looks
very different in all models. It is
\ba
\rho_\mu\left[\Phi,\partial^\mu\Phi\right]&&\hbox{\rm in model 1 and 2;}
\nonumber\\
\partial_\mu\rho_\nu\left[\partial^\mu\Phi,\partial^\nu\Phi\right]&&
\hbox{\rm in model 3 and}
\nonumber\\
\rho_{\mu\nu}\left[\partial^\mu\Phi,\partial^\nu\Phi\right]&&
\hbox{\rm in model 4.}
\ea
So one sees that even the number of derivatives in the interaction is
representation dependent. These are all on-shell equivalent.
They also become off-shell equivalent if the correct pointlike pion couplings
are included, see e.g. Ref.~\rcite{Ecker2}.

Version 3 even has no obvious vector meson dominance for the pion
charge radius. Its contribution starts only at order $p^6$.
The equivalence between the different models is obvious when we
start from an underlying quark model, see e.g. \rcite{Bijnens6} since then it
becomes a choice for the auxiliary variable.

{\bf b) Nucleons:} Here CHPT is possible for {\em some} processes.
I.e. those where the conservation of baryon number allows us to
systematically keep the heavy nucleon mass locked up inside the nucleon.
Then the pion momenta can remain small and the problem of Fig.~\tref{fig4}
does not occur.
As an example, the process $p\overline{p}\to\pi\pi$ is definitely {\em not}
treatable using CHPT but $\pi p\to \pi p$ probably is\rcite{kaiser}.
Best is to choose a formulation where the heavy mass is obviously absent
from the pion momenta. this can be done using nonrelativistic field theory for
the nucleons or using heavy baryon CHPT
(see \rcite{bernard1} and references therein).

Here there are a lot of problems and challenges.
\begin{enumerate}
\item The number of parameters is very large.
\item The mass gap between lowest states and excitations is
much smaller: $\Delta=m_\Delta-m_N\approx m_\pi$. In fact one can also
do a rigorous perturbation expansion choosing $\Delta$ as small and then
doing an expansion in $\Delta, m_\pi, \vec{p}_N, \vec{p}_\Delta$ and
$p_\pi$.
\item In the traditional view $\Delta$ is taken as large\rcite{bernard1}.
\end{enumerate}
In fact case 2 seems to follow from assumptions about leading
$1/N_c$\rcite{dashen} or about the spectrum\rcite{weinberg5}.

The field of many nucleons is also not well developed. One qualitative
conclusion is that chiral symmetry explains the observed smallness of the
3-body potential\rcite{weinberg6}

\rsection{Conclusions}
The present state of CHPT can be summarized simply. It is a mature field for
processes with mesons only, it is in its adolescent stage for calculations
involving one nucleon or one baryon and the many nucleon-baryon sector
is in its infancy.

CHPT is a useful technique despite its large number of free parameters. It is a
{\em theory}, not a {\em model}. This means that it also tells us when the
corrections are very large and its predictions thus unreliable. The
technique also allows us to use the full field theory machinery to its full
advantage.

This talk contained some discussions about the general method and some examples
of uses of CHPT. In the latter I have emphasized the work I have been
involved in.

\rsection{Acknowledgements}
I would like to thank the organizers
and their students for a pleasant and well organized meeting.


\end{document}